\documentclass[aps,pre,twocolumn,showpacs,groupedaddress]{revtex4}

\usepackage{longtable}
\usepackage{graphics}
\usepackage{amsmath}
\usepackage{amssymb}
\usepackage{amsfonts}
\usepackage{bm}
\usepackage{rotate}

\usepackage{epsfig}
\usepackage[american]{babel}
\usepackage[dvipdfm]{color}
\usepackage[latin2]{inputenc}
\usepackage[american]{babel}
\usepackage{color}
\usepackage{graphicx}

\begin{document}
\title{Equilibriumlike invaded cluster algorithm: critical exponents and dynamical properties}

\author{I. Balog}
\email{balog@ifs.hr}
\author{K. Uzelac}
\email{katarina@ifs.hr}

\affiliation{Institute of Physics, P.O.Box 304, Bijeni\v{c}ka cesta 46, HR-10001 Zagreb, Croatia}

\begin{abstract}

We present a detailed study of the Equilibriumlike invaded cluster algorithm (EIC), recently proposed as an extension of the invaded cluster (IC) algorithm, designed to drive the system to criticality while still preserving the equilibrium ensemble. We perform extensive simulations on two special cases of the Potts model and examine the precision of critical exponents by including the leading corrections. We show that both thermal and magnetic critical exponents can be obtained with high accuracy compared to the best available results. The choice of the auxiliary parameters of the algorithm is discussed in context of dynamical properties. We also discuss the relation to the Li-Sokal bound  for the dynamical exponent $z$.

\end{abstract}


\date{\today}

\maketitle

\section{Introduction\label{secintro}}

 As an important tool in the study of phase transitions Monte Carlo simulations passed through significant advances in the last decades and  number of new algorithms and improvements were proposed in various directions, such as cluster algorithms by Swendsen and Wang \cite{SwendsenWang} or Wolff \cite{Wolff}, multicanonical algorithm \cite{JK95}, Wang-Landau algorithm \cite{WL01}. An interesting direction undertaken by Machta et al. \cite{MachtaChayes} was to construct an algorithm which would, in a self organized way, drive the system to critical temperature without prior knowledge of it. It was applied to various classical models from discrete to continuous ones including some more complicated effects such as frustration \cite{FCC98}, quasi periodic ordering \cite{RB00}, or tricritical points \cite{BU07}. Their method, based on the invasion percolation \cite{WW} and the cluster algorithm, appeared to converge much faster than in the Swendsen-Wang (SW) algorithm, with a dynamical exponent $z$ close to zero. In the same time it opened a number of issues \cite{Moriarty},\cite{GLL05} related to the fact that the underlying nonequilibrium procedure does not generate an equilibrium ensemble and consequently is not able to reproduce correct critical exponents when a finite size scaling is applied. There were also some other attempts to design a cluster algorithm that would selfregulate to criticality \cite{TomitaOkabe,Fulco00}. 

Recently we proposed a modification of the IC algorithm  \cite{eic} aimed to reestablish the equilibrium in the self-regulating procedure of the IC algorithm. The principal idea is to impose a simple constraint on the temperature uncertainty characteristic to the IC algorithm, reducing it to the limits compatible with the equilibrium distribution. As shown in our preliminary work \cite{eic}, when this single constraint is applied to the IC algorithm, the correct scaling properties of thermodynamic observables are recovered, while the algorithm still retains the property of self driving to criticality.

The aim of the present paper is twofold. First, it offers a more intensive and comprehensive numerical calculations, including the leading convergence exponent, in order to examine the precision of the results for critical exponents and temperature that the EIC algorithm can provide. Second is to extend the study to the dynamical properties of the algorithm, in particular to the autocorrelation functions, which indicate deviations from the equilibrium distribution providing the criteria for the choice of auxiliary parameters of the method, which would not produce systematic errors. 

The paper is organized as follows. In Sect. \ref{meth} we introduce the model and notations and describe in detail the principles of EIC algorithm. In Sect. \ref{crit} is presented the analysis of critical exponents and temperature including the leading correction to scaling. We also add a brief comment on the application of this algorithm to the first-order phase transition. In Sect. \ref{dyn} we study the dynamical properties of EIC algorithm through the autocorrelations of energy and order parameter with a special emphasis on the impact of choice of the two auxiliary parameters. Sect. \ref{discs} contains the conclusion.

\section{Algorithm\label{meth}}

 We consider the Potts model \cite{Potts,Wu} defined by the Hamiltonian
\begin{equation}
\label{dil_Potts}
H=-J\sum_{<i,j>}\big(\delta_{\sigma_i,\sigma_j}-1\big),
\end{equation}
where $\sigma_i$ denotes the $q$-state Potts variable at the lattice site $i$ and the summation runs over the nearest neighbors. As shown by Fortuin and Kasteleyn (FK) \cite{FortuinKasteleyn}, the partition function of the model (\ref{dil_Potts}) 
can be written as
\begin{equation}
\label{z-potts}
Z = \sum_{\{\sigma_i\}} e^{-\beta H}  \equiv  \sum_{\{\sigma_i\}}\prod_{<i,j>} 
\big( (1 - p) + p \cdot \delta_{\sigma_i\sigma_j}\big), 
\end{equation}
where 
\begin{equation}
\label{p}
p=1-e^{-\beta J}, 
\end{equation}
and $\beta$ is the Boltzmann factor.
By expanding the binomial product into sum over graphs and integrating over the spin degrees of freedom it reduces to the random-cluster (RC) model 
\begin{equation}
\label{rand_cluster}
Z=\sum_{\gamma \in\Gamma}p^{b(\gamma)} \cdot (1-p)^{B-b(\gamma)} \cdot q^{c(\gamma)},
\end{equation}
which can be understood as a generalized bond percolation, where $p$ is the bond probability. The summation in Eq. (\ref{rand_cluster}) is taken over the set $\Gamma$ of all the graphs on the lattice, each graph $\gamma$ representing one possible bond configuration. $B$ is the number of the lattice edges, $b(\gamma)$ denotes the number of bonds, and $c(\gamma)$ is the number of connected components (FK clusters) in the graph $\gamma$.

FK expansion of the Potts model partition function is also used as a basis for construction of cluster algorithms for numerical simulations \cite{SwendsenWang}, \cite{ES88}. To this purpose the r.h.s. of the Eq. (\ref{z-potts}) can be written in alternative way by introducing new degrees of freedom $n_{i,j}=0,1$, representing the absence or presence of bonds in the expansion of the binary product in Eq. (\ref{z-potts}). In such a way the Boltzmann weight in (\ref{z-potts}) is decomposed into two joint distributions: over spin  ($\{\sigma\}$) or bond ($\{n\}$) configurations
\begin{equation}
\label{dist-sw}
w(\{\sigma\},\{n\})  = \frac{1}{Z}  \cdot \prod_{<i,j>} \big ( (1-p) \cdot \delta_{n_{i,j},0} + p \cdot \delta_{n_{i,j},1} \cdot \delta_{\sigma_i,\sigma_j} \big).
\end{equation}
As it may be seen from Eqs (\ref{z-potts}) and (\ref{dist-sw}), for a given bond configuration, only the spin configurations where the spins belonging to the same cluster are in the same state give a nonzero contribution to the partition function, while entire clusters may be flipped without an energy cost. On the other hand, for a given spin configuration ($\{\sigma\}$) the Boltzmann weight reduces to a binomial distribution of bond occupancy
\begin{equation}
\label{fixed-s}
w\big(\{\sigma\}\big)  = \frac{1}{Z} \cdot (1-p)^{B-n_{ss}} \cdot \sum_{b} {n_{ss}\choose b} ~ p^b \cdot (1-p)^{n_{ss}-b},
\end{equation}
where $n_{ss}$ denotes the total number of "satisfied" edges (surrounded by spins in the same state). The  standard cluster algorithms such as SW \cite{SwendsenWang} consist in alternate updates of bonds and spins. In the first part of the MC step, taken a given spin configuration $\{\sigma\}$, the FK clusters are formed by putting bonds, with probability $p$, between neighbors in the same state. In the second part each cluster is flipped at random producing a new configuration of spins. The bonds are then erased and one proceeds to the next MC step. In such a way the equilibrium ensemble is generated.

The IC algorithm \cite{MachtaChayes} consists of the same two steps, but the bond update is altered in order to accomplish the self driving to criticality. Given the configuration of spins, the bonds are added with the probability 1, but only until the percolation is achieved. This is interpreted as a signature of a finite-size critical point. As soon as the percolation is reached the FK clusters are randomized. The fraction of "satisfied" edges that has been populated 
\begin{equation}
\label{p_gamma}
p_{\{\sigma\}}=\frac{b(\{\sigma\})}{n_{ss}(\{\sigma\})},
\end{equation}
is recorded after each MC step. $b(\{\sigma\})$ denotes the number of added bonds and $n_{ss}(\{\sigma\})$ the number of neighboring pairs in a same state for a given spin configuration $\{\sigma\}$. The average over simulations $\overline{p}_{\{\sigma\}}$ is identified with the quantity $p$ defined in (\ref{p}), yielding a critical temperature estimate. It is easy to understand why such a procedure is self-driving to criticality. Namely if the given configuration of spins is typical for a temperature much higher than $T_c$ (with random and poorly correlated spins) almost all the edges with satisfied neighbors will need to be populated before the percolation has been achieved. The $n_{ss}$ in the next iteration can only be larger - the system will be driven towards configurations with lower energies, corresponding to lower temperatures. On the other hand, when the configuration is typical of lower temperatures, with $n_{ss}$ considerably larger than the one in criticality, small amount of added bonds will suffice to achieve percolation, so that, after randomization of clusters $n_{ss}$ will be diminished driving the system to higher temperature configurations. In the same time one can notice here a back-draw of this self driving procedure. In both cases, the described configurations that have been retained for statistics result from the tails of a binomial distribution (\ref{fixed-s}). Problem is that such configurations are typical during entire IC simulation. Namely, in spite of the fact that the system is driven to criticality, the described oscillations remain significant throughout the simulations, so that it generates distribution of $p_{\{\sigma\}}$ that is much wider than it would result from the binomial distribution of (\ref{fixed-s}). Consequently, the resulting ensemble is not a canonical one and not in equilibrium, as observed already in \cite{Moriarty}.  

Our EIC algorithm approach follows the same IC procedure but restrains the excessive fluctuations in $p_{\{\sigma\}}$ in order to regain the equilibrium fluctuations. The main idea is to constrain the width of distribution of $p_{\{\sigma\}}$ obtained from the simulations by the relation (\ref{p_gamma}) to the width which is compatible with the distribution of $b_{\{\sigma\}}$ in (\ref{fixed-s}) and which has the same scaling in $L$. The width of the binomial distribution in Eq. (\ref{fixed-s}) gives $var(p_{\{\sigma\}}) \propto \sqrt{p(1-p)} n_{ss}^{-d/2}$. Since $n_{ss}$ corresponds to the energy of a given configuration and is $\propto L^{d}$, we impose the constraint on $p_{\{\sigma\}}$, which allows it to vary only within limited range of the width $v \propto L^{-d/2}$. 

To include this constraint into algorithm, we modified the stopping rule by introducing two auxiliary parameters to the IC algorithm,  $\tilde{v}$ and $N_a$. Let us review this procedure briefly and discuss the role of these free parameters.

The MC iterations are grouped in intervals of $N_a$ MC steps. In every interval $i$ of $N_a$ steps the bonds are placed on the lattice between satisfied neighbors with the intention to construct a percolating cluster. However, the system is not left to percolate with just any value of $p_{\{\sigma\}}$ as in the IC algorithm. Any accepted configuration is required to lie inside the interval: $\overline{p}_{i-1}-v<p_{\{\sigma\}}< \overline{p}_{i-1}+v$, where $\overline{p}_i$ denotes the average of $p_{\{\sigma\}}$ over the i-th block of $N_a$ steps. Here $v$ is the free parameter set to be: 
\begin{equation}
\label{defv}
v=\tilde{v}\cdot L^{-d/2} 
\end{equation}  
 and $\tilde{v}$ is a constant whose choice will be discussed later. If the system percolates before the lower bound of  $p_{\{\sigma\}}$ is reached, the bonds are still added until the lower bound is attained. If the upper bound of $p_{\{\sigma\}}$ is reached before the percolation is attained, the process is stopped without requiring the system to percolate. After every block of $N_a$ MC steps, the new average, $\overline{p}_{i}$, is calculated to be used as a reference value in the next $N_a$ steps. Obviously, in this way the information on the value of $\overline{p}_{i-1}$ in the preceding $N_a$ MC steps is propagated to the consecutive interval $i$ and will have an effect on the autocorrelation functions (to be discussed in the Sect. \ref{dyn}). 

The starting value $\overline{p}_{0}$ can be obtained by the unconstrained IC algorithm. 
One can also define a faster equilibration of the IC algorithm by gradually reducing the value of parameter $\tilde{v}$ in (\ref{defv}) during some initial set of intervals of $N_a$ steps. These first few blocks of $N_a$ steps are discarded from statistics and the configurations are recorded after the steady state was reached. 

Distribution of $p_{\{\sigma\}}$ resulting from the above procedure has the required width proportional to $v=\tilde{v}\cdot L^{-\frac{d}{2}}$ under the assumption that $N_a$ is large enough so that the fluctuations of $p_{\{\sigma\}}$ within the interval are dominant over the fluctuations of $\overline{p}_{i}$ between the intervals. In this case the fluctuations of the mean value $\overline{p}_{i}$ reduce to the actual temperature fluctuations, producing the width of order $L^{-1/\nu}<L^{-d/2}$ if the heat capacity exponent $\alpha>0$, or of order $L^{-d/2}$ if $\alpha<0$. If $N_a$ would be small enough for the fluctuations of $\overline{p}_{i}$ between the intervals to take over, a crossover to the IC regime would be seen. Later we will demonstrate that there exists a clear indication when this occurs.
\begin{figure}[!hbt]
\includegraphics[scale=0.9]{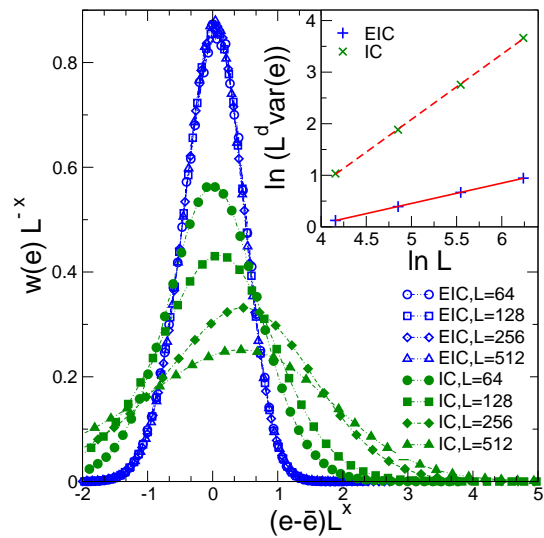}
\caption{(Color online) Overlap fits of energy distributions $w(e)$ for $q=3$ Potts in 2D rescaled with the exact exponent $X=0.8$. In the inset are shown ln-ln plots of $L^d\cdot var(e)$ vs. $L$. The statistic is $10^6$ MCS for both EIC and IC cases, and the parameters of the EIC algorithm were chosen to be $N_a=100$ and $\tilde{v}=\frac{1}{10}$.}
\label{fig1}
\end{figure} 

In Fig. \ref{fig1} we give an illustration that the canonical form of energy distribution is recovered within the EIC algoritm. As noted by Machta etal. \cite{MachtaChayes}, $L^d\cdot var(e)$ for 2D Ising obtained by the IC algorithm scales approximately linearly with the system size $L$, while it should scale as $ln L$ if the ensemble was canonical. To examine the same point within the EIC algorithm, we present in Fig. \ref{fig1} an overlap fit of the energy distributions resulting from EIC algorithm and compare it with the distributions obtained by IC algorithm with equal statistics, for several lattice sizes in the case of $q=3$ Potts in 2D. The distributions are rescaled with the exact value of the exponent describing the width of the energy distribution in the equilibrium ensemble, $X=\frac{d}{2}-\frac{\alpha}{2\nu}=0.8$. It is immediately seen that the results for different $L$, obtained by EIC, collapse very well, while the ones obtained by IC do not. To quantify that fact, we show in the inset of Fig. \ref{fig1} the ln-ln plots of $L^d \cdot var(e)$ vs. $L$, the slopes of which in the canonical ensemble represent the exponent ${\alpha}/{\nu}$ (exactly equal to $0.4$ in this case). For the lattice sizes considered we obtain $\approx0.38$ and $\approx1.26$ for EIC and IC cases respectively (in terms of the width scaling  $X_{EIC}\approx0.81$ and $X_{IC}\approx0.37$). This indicates that the ensemble sampled by the EIC algorithm is indeed canonical, contrary to the one produced by the standard IC algorithm, which appears to be much wider and scales with different exponent, characterizing the algorithm itself \cite{Moriarty}. As it shall be seen in the next section, the EIC algorithm can be used to obtain very accurate values for critical exponents, not only the magnetic one related to criticality, but also the thermal critical exponent related to the approach to criticality and implying the validity of fluctuation-dissipation theorem.  


\section{Critical behavior\label{crit}}

Derivation of critical exponents and temperature by EIC algorithm was already illustrated by applying it to several cases of the Potts model in our preliminary study. In present work, we selected only two cases both exhibiting 2nd order phase transition, one in two and one in three dimensions: case $q=3$ in 2D and Ising case ($q=2$) in 3D. We conducted extensive numerical simulations increased by an order of magnitude with respect to the preliminary study \cite{eic}. This allowed us to calculate the leading power-law corrections as well and analyze the precision of the results with particular attention to possible systematic errors due to the choice of the auxiliary parameters. Unlike our previous study, we avoided to use any data exactly known in advance and applied consistently multi parameter fits, which give critical temperature and critical exponents simultaneously whenever needed.

   The simulations were performed on lattices with periodic boundary conditions and sizes up to $L=96$ and $L=1024$ in the 3D and 2D case respectively. The statistics varied from $15 \cdot 10^6$ iterations for smaller to $7\cdot 10^6$ iterations for the largest lattice sizes in the 3D case and from $15\cdot 10^6$ to $5\cdot 10^6$ in the 2D case. The auxiliary parameters were set to $\tilde{v}={1}/{10}$ and $N_a=100$ in both cases. The percolation was established by the topological rule, i.e. by the condition that the infinite cluster wraps around the lattice. The time necessary to reach the equilibrium state never exceeded the first 3000 MC steps with the above choice of auxiliary parameters $N_a$ and $v$. 

The running time per MC step was approximately the same as for the original IC algorithm. For illustration, a run of $10^5$ MC steps for the 2D Ising model on $L=64$ lattice requires approximately 500s on the AMD Opteron 240 processor (1.4GHz).

\subsection{The magnetic critical exponent $y_h$}
 
The magnetic critical exponent was evaluated independently from three different quantities calculated at criticality, the average mass of the largest cluster, the order parameter and the susceptibility.

The average mass of the largest cluster $\overline{s}_{max}$ is defined as the number of spins in the largest cluster averaged over all the configurations. Its scaling yields the critical exponent $y_h$ directly 
\begin{equation}
\label{smax}
\overline{s}_{max}|_{p=p_c} = a_s\cdot L^{y_h}+b_s\cdot L^{y_h+\omega_1}+\cdots.
\end{equation}
\begin{figure}[!hbt]
\includegraphics[scale=0.9]{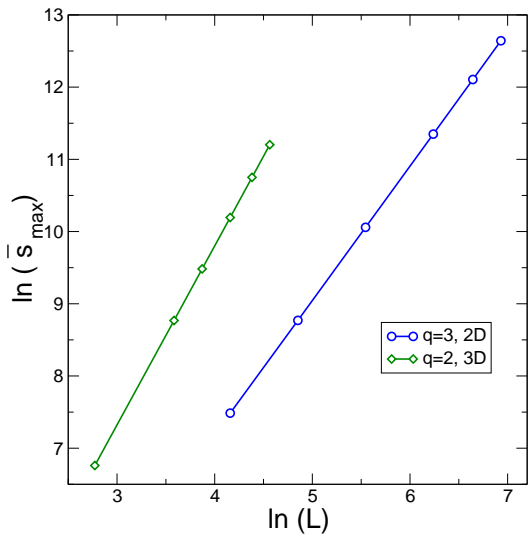}
\caption{(Color online) Largest cluster mass $\overline{s}_{max}$ versus the system size.}
\label{fig2}
\end{figure} 
A four-parameter fit of our finite-size results (see Fig. \ref{fig2}) to the form  (\ref{smax}) produced rather accurate estimate of  the exponent $y_h$ and showed that the leading correction is of a power law form. Results for the exponents are: $y_{h}=2.4815(5)$ with $\omega_{1}=-2.2(1)$ and $y_{h}=1.8661(7)$ with $\omega_{1}=-1.1(1)$ for 3D Ising and 2D q=3 Potts model respectively. 

The order parameter $m$ scales at $T_c$ according to
\begin{equation}
\label{mscl}
{m}|_{p=p_c}= a_m\cdot L^{-\beta/\nu}+b_m\cdot L^{-\beta/\nu+\omega_2}+\cdots,
\end{equation}
where $\beta/\nu$ is related to $y_h$ by $\beta/\nu=d-y_h$. It was calculated by using the standard definition of the order parameter for the Potts model 
\begin{equation}
\label{m}
m = \overline{m}_1, \hskip 20pt   m_1 = \frac{q}{(q-1)\;L^d}\; max_{\alpha} \left[ \sum_{i}\; \left(\delta_{\sigma_i,\alpha} - \frac{1}{q} \right) \right]. \\
\end{equation}
which consists in taking the most populated among the $q$ Potts states in each configuration. By a four-parameter fit to the form  (\ref{mscl}) we found the exponents $\beta/\nu =0.5180(2)$ with $\omega_{2}=-1.9(1)$ and $\beta/\nu =0.1345(5)$ with $\omega_{2}=-2.2(1)$ for 3D Ising and q=3 case in 2D respectively. 

The fluctuations of the magnetization in equilibrium are related to the susceptibility by the relation
\begin{equation}
\label{chi}
\chi = \; \frac{L^d}{k_B T} \; \left[\overline{m^2}-\overline{m}^2 \right],
\end{equation}
which assumes the validity of the fluctuation-dissipation theorem, not fulfilled for the standard IC algorithm. 
\begin{figure}[!hbt]
\vskip45pt
\includegraphics[scale=0.9]{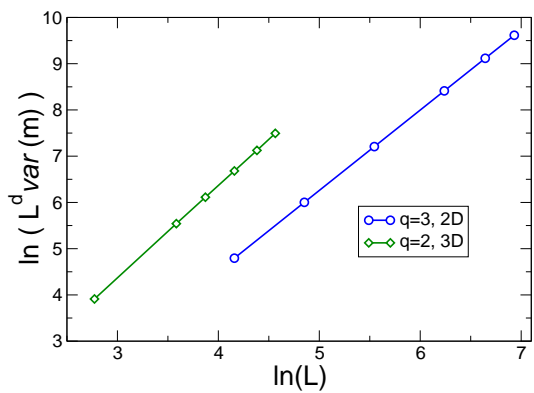}
\caption{(Color online) Order parameter fluctuations plotted versus the system size $L$.}
\label{fig3}
\end{figure}
For a finite system, at the quasicritical point $T_{cL}$, the susceptibility exhibits maximum, which scales with a power-law \begin{equation}
\label{chi0}
\chi|_{p=p_{cL}}\propto L^{\gamma/\nu},
\end{equation}
and is related to the magnetic exponent by $\gamma/\nu=2\cdot y_h-d$. The log-log fit of the results (see Fig \ref{fig3}) gives $\gamma/\nu =1.98(1)$ and $\gamma/\nu =1.74(1)$ for 3D Ising and q=3 case in 2D respectively. We find the values of $\gamma/\nu$ less accurate than the estimates for $y_h$ or $\beta/\nu$. The reason may be attributed to the fact that, for a finite system, the critical point identified by the percolation through a specific choice of the stopping rule, should not necessarily coincide with the quasi-critical temperature defined by the maximum of the susceptibility.

Another quantity of interest, related to the magnetization is the Binder fourth order cumulant \cite{Binder}, defined as
\begin{equation}
\label{U4}
U_4=1-\frac{\overline{m^4}}{3\cdot\overline{m^2}^2}.
\end{equation}   
One of its main benefits within numerical calculations is to provide a good estimation of the critical temperature, which in the present case, where the algorithm drives itself to the critical point is not the main issue. At criticality it may be related to the one of the universal amplitude ratios, and we found interesting to examine this quantity and its size convergence within the present algorithm.

\begin{figure}[!hbt]
\includegraphics[scale=0.9]{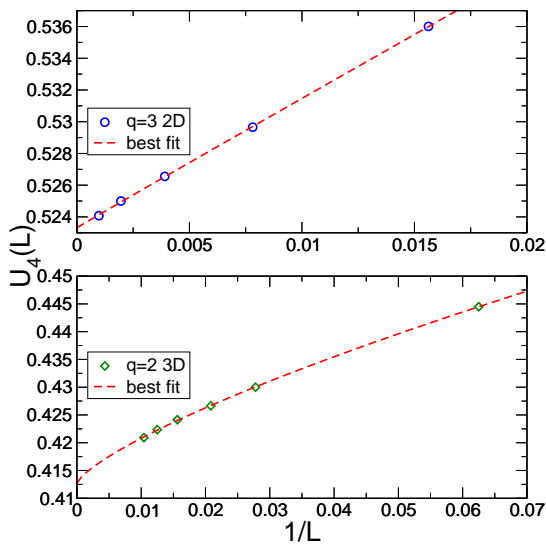}
\caption{(Color online) EIC data for Binder cumulant $U_4(L)$  versus $1/L$.}
\label{fig4}
\end{figure}

It is important to notice, that in the present approach, $U_4$ is calculated at the finite size critical point defined by the topological rule, hence $\tau^{\nu} L = \tilde{a}$, where $\tilde{a}$ is some non universal constant factor. The $L$ dependence of Binder cumulant at this finite size critical point is
\begin{equation}
\label{U4L}
U_4(L,t,u)=U_4(\tau^{\nu} L)+b_1uL^{y_i}+\cdots.
\end{equation}   
Consequently, the constant term of the Binder cumulant in the thermodynamic limit depends on the way of approaching to the critical point. Other authors \cite{Hasenbusch} considered instead the universal value $U_4(0)$. In Fig \ref{fig4} 
the results are shown for 3D Ising and 2D $q=3$ cases. The values can be fitted quite accurately to a three-parameter power law form: $U_4(L)=U_4(\tilde{a})+b_{U,1}\cdot L^{\omega_3}$. The constants $U_4(\tilde{a})$ are found to be $U_4(\tilde{a})=0.412(4)$ and $U_4(\tilde{a})=0.523(2)$ for 3D Ising and q=3 Potts in 2D respectively. The respective convergence exponent $\omega_3$ is determined to be $\omega_3=-0.75(3)$ and $\omega_3=-0.91(4)$. A value $U_4(L\to\infty)\approx 0.40$ which is similar to the value obtained in the present work for 3D Ising case can be found in Ref. \cite{Ferrenberg} which was obtained by the extrapolation of th finite-size values $U_4(L)$ taken at the point where $\frac{\partial U_4(L)}{\partial J}$ attains its maximum.

\subsection{The thermal critical exponent $y_t$ and critical temperature}

   Two quantities related to the thermal critical exponent $y_\tau=1/\nu$ were  examined. The first one is $p_c(L)$, related to the quasicritical temperature, with the leading size dependence
\begin{equation}
\label{pcl}
p_c(L)=p_c(L\rightarrow\infty)+a_p\cdot L^{-1/\nu}+\cdots.
\end{equation}
It is obtained from the average over $p_{\{\sigma\}}$ defined in the Eq. (\ref{p_gamma}).
\begin{figure}[!hbt]
\includegraphics[scale=0.9]{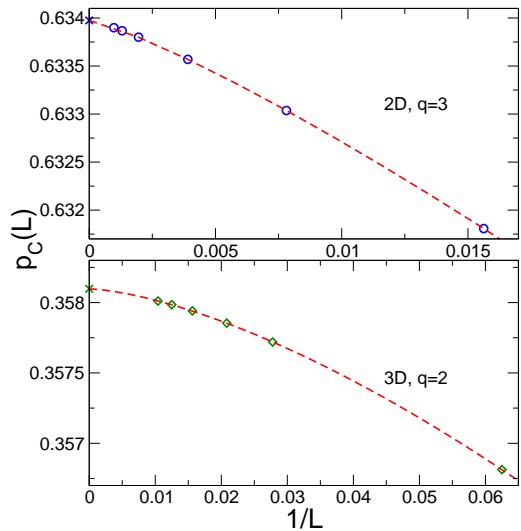}
\caption{(Color online) Data for $p_c(L)$ versus $1/L$. Dotted lines describe three parameter fits and crosses denote exact, or best known values.}
\label{fig5}
\end{figure} 
Data for $p_c(L)$ with a three-parameter fit are illustrated in Figure  \ref{fig5}. The fit appears to be more 
 suitable for finding the $p_c(L\rightarrow\infty)$, giving the critical point with the precision up to the sixth digit, yielding $p_c(L\rightarrow\infty)=0.358097(1)$ and $p_c(L\rightarrow\infty)=0.633975(1)$ for 3D Ising and q=3 case in 2D respectively.   An estimate of the exponent $y_t$ by the same procedure appears, however, to be by an order of magnitude less precise than by the logarithmic derivative of $m$.

\begin{figure}[!hbt]
\includegraphics[scale=0.9]{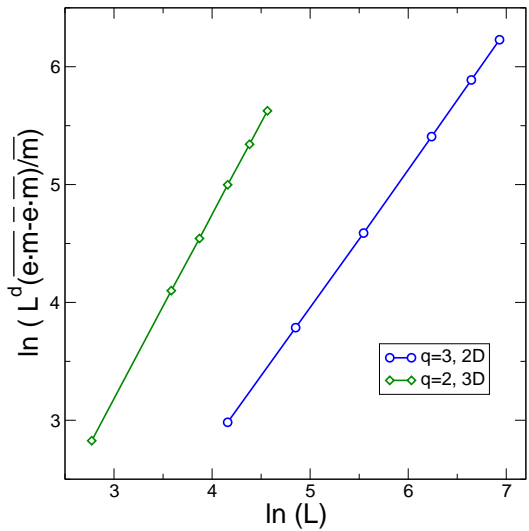}
\caption{(Color online) Logarithmic derivative of magnetization $\frac{\partial \ln m}{\partial \beta}$ plotted versus the system size.}
\label{fig6}
\end{figure}

To derive the critical exponent $\nu$ we consider the logarithmic derivative of magnetization (see Fig. \ref{fig6}) and use 
\begin{equation}
\label{em}
\frac{\partial \ln m}{\partial \beta}=L^{d}\cdot\frac{\overline{em}-\overline{e}\cdot\overline{m}}{\overline{m}}\propto a_{em}\cdot L^{1/\nu}+b_{em}\cdot L^{y_t+\omega_4}+\cdots.
\end{equation}
The equality in Eq. (\ref{em}) implies that the equilibrium ensemble is well reproduced by our algorithm, which we expect to hold. Indeed, the obtained values for the exponent $\nu$ are in excellent agreement with known exact, or best approximate results. The four-parameter fit to the form of the r.h.s. in (\ref{em}) gave $y_{\tau}=1.586(5)$, $\omega_{4}=-2.0(2)$ and $y_{\tau}=1.201(8)$, $\omega_{4}=-1.0(3)$ for 3D Ising and 2D, q=3 case respectively.  

\subsection{Summary of critical exponents and corrections to scaling}

\begin{table}[!h]
\label{tab1}
\begin{ruledtabular}
\caption{Leading critical parameters obtained by EIC (from $L=64$ to $1024$ for $q=3$ 2D and from $L=16$ to $96$ for 3D) compared to best known or exact results.}
\centering{
\begin{tabular*}{8.0cm}{c@{}c@{}c@{}c@{}c@{}}
 & \multicolumn{2}{c}{$q=2$, 3D} & \multicolumn{2}{c}{$q=3$, 2D}\\
\hline
 & EIC & best known\footnotemark[1] & EIC & exact$^*$ \\
\hline
\hline
$p_c(L\rightarrow\infty)$ & $0.358097(1)$ & $0.358098(3)$ & $0.633975(1)$ & $0.6339745\dots$ \\
$y_{\tau}$ & $1.586(5)$ & $1.587(6)$ & $1.201(8)$ & $\frac{6}{5}$ \\
$y_h$ & $2.4815(5)$ & $2.4818(6)$ & $1.8661(7)$ & $1.8\dot{6}$ \\
$\beta/\nu$ & $0.5180(4)$ & $0.5182(6)$ & $0.134(1)$ & $0.1\dot{3}$ \\
$\gamma/\nu$ & $1.98(1)$ & $1.964(1)$ & $1.74(1)$ & $1.7\dot{3}$ \\
\end{tabular*}
}\end{ruledtabular}
\footnotetext[1]{Best known values from \cite{Pelissetto} and \cite{Ferrenberg}.}
\end{table}

\begin{table}[!h]
\label{tab2}
\begin{ruledtabular}
\caption{Corrections to scaling obtained by EIC algorithm (for reference see in the text)}
\centering{
\begin{tabular*}{8.0cm}{c@{}c@{}c@{}}
 & $q=2$, 3D & $q=3$, 2D\\
\hline
\hline
$\omega_1$ & $-2.2(1)$ & $-1.1(1)$ \\
$\omega_2$ & $-1.9(1)$ & $-2.2(1)$ \\
$\omega_3$ & $-0.75(3)$ & $-0.91(4)$ \\
$\omega_4$ & $-2.0(2)$ & $-1.0(3)$ \\
\end{tabular*}
}\end{ruledtabular}
\end{table}

Critical exponents, and their leading corrections calculated in preceding two sections are summarized in Tables I and II. It can be concluded that, with only a fraction of numerical effort needed for the SW algorithm, the critical behavior can be obtained very accurately and in excellent agreement with the best known or exact results. The location of the critical point can be obtained with rather high degree of accuracy as well. 

For better reliability of our results we included in our analysis the corrections to scaling.
We observe that they differ substantially from the expected scaling corrections due to the irrelevant scaling fields and nonlinearities of scaling fields near criticality and should be attributed to the method itself. The leading irrelevant exponent for 2D $q=3$ is exactly known to be $y_i=-\frac{4}{5}$. 
For the 3D Ising case it is estimated to be $y_i=-0.84(4)$ \cite{Pelissetto}. Our results show the scaling corrections that decay much faster than $y_i$ in most cases, giving the exponents $\omega_i$ close to -2 or -1 in the $3D$ and $2D$ case respectively. The exception is the convergence of the Binder cumulant ($\omega_3$ in Tab. II) with the correction exponent that can be compared to the value of leading irrelevant scaling field $y_i$ found by other authors. 

\subsection{First-order phase transition}

The example for our study being the Potts model gives and opportunity to examine the applicability of our algorithm to the first order phase transitions. We make thus a small digression here, to show, without going into any details, that the same procedure, when applied to a first order phase transition gives a clear indication on its character.

\begin{figure}[!hbt]
\includegraphics[scale=0.8]{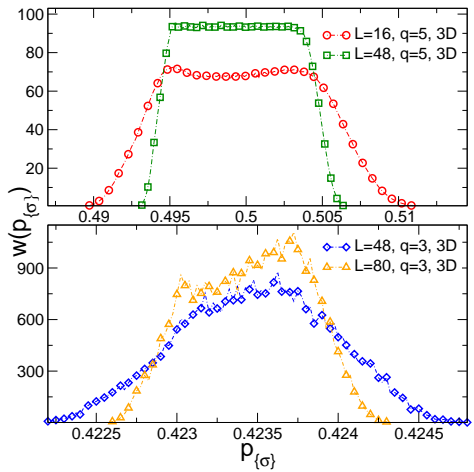}
\caption{(Color online) Distributions $w(p_{\{\sigma\}})$ of $p_{\{\sigma\}}$ produced by EIC algorithm for the 3D Potts model in the case of strong ($q=5$) and weak ($q=3$) first-order phase transition.}
\label{fig7}
\end{figure}

We first remind that for studies of the first-order transitions by the IC algorithm the stopping rules based on the cluster mass are more appropriate \cite{MachtaChayes} and can be readily extended to EIC. 
Our intention here is to point out how the first-order phase transition manifests when the topological stopping rule is used. As the first-order transition is characterized by a finite hysteresis width, the construction of percolating clusters becomes possible for all the values of $p_{\{\sigma\}}$ within the hysteresis and thus the values of $p_{\{\sigma\}}$ sweep trough the entire width of the hysteresis. 
This results with the plateau in the distribution $w(p_{\{\sigma\}})$ who's width is not proportional to $L^{-\frac{d}{2}}$ but tends to some constant value corresponding to the hysteresis width in the thermodynamic limit (see Fig. \ref{fig7}). 
We illustrate this on the two examples of the first-order transition in the 3D Potts model. For $q=5$, the first-order transition is very strong and the figure clearly reveals the width of the hysteresis. For $q=3$ which is characterized by a very weak first-order transition, larger lattices are needed to see the hysteresis. However, already for modest sizes, an inspection of the Figure \ref{fig7} suggests that the width of the distribution will not vanish with some power-law in $L$, since its shape starts to qualitatively resemble the one for 3D, $q=5$.

\section{Dynamical properties of EIC algorithm\label{dyn}}

The dynamics introduced by the EIC algorithm was studied by considering the autocorrelation function defined for an arbitrary observable $O$ by
\begin{equation}
\label{ac_def}
\Gamma_{O}(t)=\frac{\overline{O(t)O(0)}-\overline{O}^2}{\overline{O^2}-\overline{O}^2}.
\end{equation}
The average is taken over the ensemble generated in the single run, after the thermalization, i.e. 
\begin{eqnarray}
\label{acorr_mc}
\overline{O(t')O(0)} &=& \frac{1}{t_{max}-t'} \sum_{t=1}^{t_{max}-t'} O(t) O(t+t'),\\
\overline{O^n} &=& \frac{1}{t_{max}} \sum_{t=1}^{t_{max}} O(t)^n, 
\end{eqnarray}
where $t$, and $t'$ denote time in discrete integer units corresponding to one MC step, and counting starts after equilibration. $t_{max}$ denotes the number of counted steps in the single run. The Eq. (\ref{acorr_mc}) implies certain equilibrium properties of macroscopic variables such as the translational invariance in time and well defined temperature (or $p$). The self-regulating procedure of EIC contains deviations from these properties, but as these occur within bounds which we expect to preserve the equilibrium ensemble we find justified to attempt a study of this process as the equilibrium one. We skip here the analysis of the nonequilibrium details of the process, which would require a different averaging procedure. Instead, we apply the Eq. (\ref{acorr_mc}) and just comment the differences found in comparison to the form of autocorrelation functions obtained in standard MC approaches. 

   In standard MC approaches one may usually distinguish three regimes of behavior of the autocorrelation functions \cite{LandauBinder}. For short times they behave as a sum of exponential decays. For longer  times they adopt a single exponential decay of the form $\Gamma_O(t)\propto e^{-\frac{t}{\tau}}$ with a correlation time $\tau$, while for very large times the correlations vanish and the statistical errors take over. 

\begin{figure}[!hbt]
\includegraphics[scale=0.9]{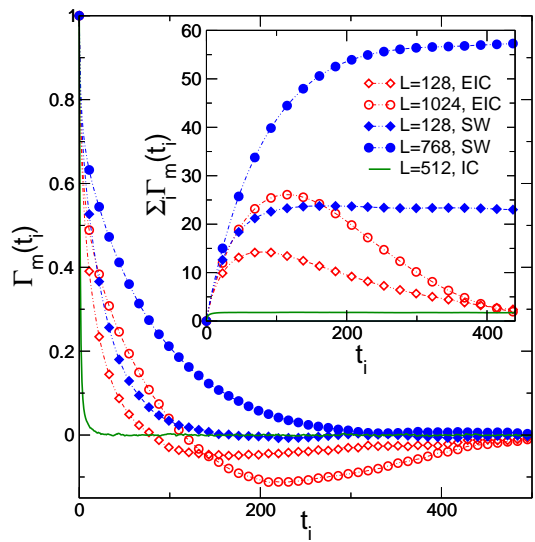}
\caption{(Color online) Magnetic autocorrelation function $\Gamma_m(t)$ for EIC, SW, and IC algorithms for 2D, q=3. On the inset of the figure are shown the integrated autocorrelation functions for the same cases.}
\label{fig8}
\end{figure}

In Fig. \ref{fig8}
 are displayed the autocorrelations in the EIC algorithm compared to those in the SW and IC algorithms on the example of the magnetic autocorrelation function in the case 2D $q=3$. While the IC autocorrelations fall extremely fast, the EIC correlations display similar decay as in the SW algorithm, with the difference that they acquire a negative sign at times of the order of $N_a$. The slower decay of $\Gamma_{O}(t_j)$ reflects the fact that IC dynamics has been slowed down by the requirement that the values of $p_{\{\sigma\}}$ remain during $N_a$ MC steps in the narrow interval of $2\cdot v$ around the average of $\overline{p}_{i-1}$.
The fact that $\Gamma(t_j)$ acquires negative values can be related to the abrupt switch of the mean value $\overline{p}_{i}$ after every $N_a$ steps as a part of the self-regulating procedure. EIC algorithm has a tendency to "overshoot" the critical temperature and find itself on the other side of the critical point in the following step. While $\overline{O}$ represents the average of the magnetization taken over the entire run, the average of magnetization over each single $N_a$ block is different (larger of smaller than $\overline{O}$) and varies according to $\overline{p}_{i}$. This explains why in the expression for the correlation function, $(O(t)-\overline{O})$ and $(O(t+t')-\overline{O})$ can in average be of the opposite sign, when $t$ and $t+t'$ belong to the neighboring blocks of $N_a$ steps and produce anticorrelations.

\begin{figure}[!hbt]
\includegraphics[scale=0.9]{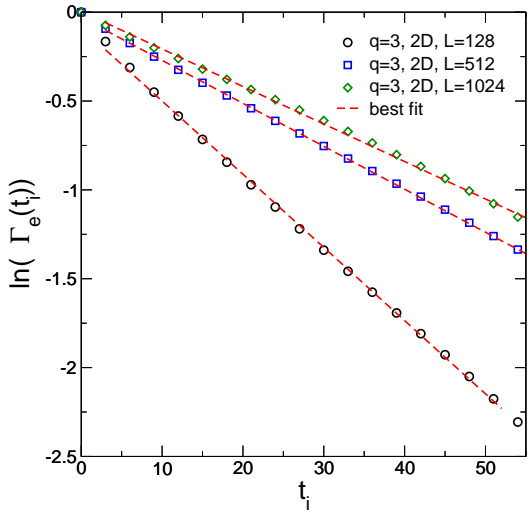}
\caption{(Color online) Energy autocorrelation functions $\Gamma_e(t)$ of $q=3$, 2D case for several lattice sizes in the exponential region. Dashed lines represent best fits.}
\label{fig9}
\end{figure}

The semi-log plot displayed in Figure \ref{fig9} shows that the autocorrelation functions indeed match well the assumed exponential form.
This leads us to the question of the correlation time and the corresponding dynamic exponent $z$, which should satisfy the scaling relation 
\begin{equation}
\label{tauz}
\tau\propto L^{z}.
\end{equation}
Derivation of the correlation time is performed usually in two equivalent ways: by fitting the exponential decay (exponential correlation time), or by taking the integral $\tau = \int_0^\infty\Gamma_O(t)dt$ (integrated correlation time).

\begin{figure}[!hbt]
\includegraphics[scale=0.9]{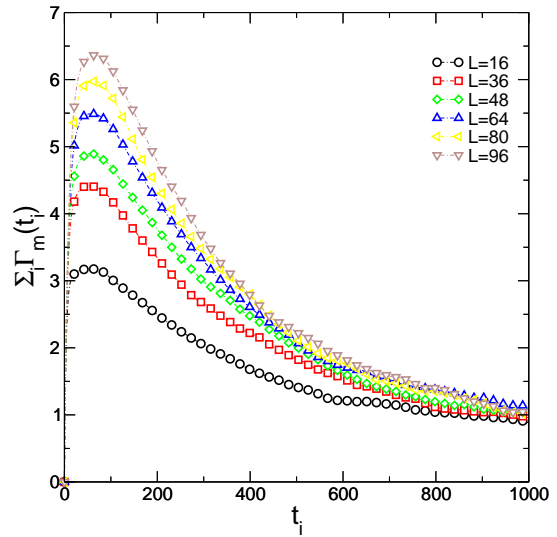}
\caption{(Color online) Integrals of magnetization autocorrelation functions $I_m(t_k)=\sum_{t_k}^{t_j=0}\Gamma_m(t_j)$ of 3D Ising for several lattice sizes.}
\label{fig10}
\end{figure}

\begin{figure}[!hbt]
\includegraphics[scale=0.9]{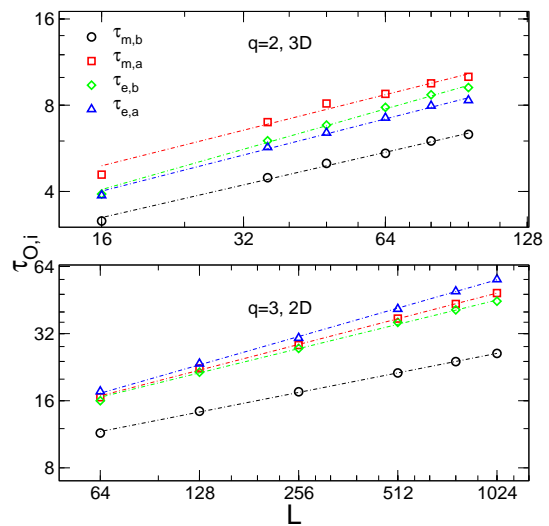}
\caption{(Color online) Characteristic times of energy and magnetization autocorrelation functions for the systems considered, determined by two approaches described in the text.}
\label{fig11}
\end{figure}

The calculation of the exponential correlation time applies to the present problem in a straightforward way, as already illustrated in Fig (\ref{fig9}). The correlation times derived from the slopes of such semi-log plots and the corresponding dynamic exponents are presented in Table III, denoted by $\tau_{O,a}$ and $z_{O,a}$ respectively. The calculation of an integrated correlation time is obviously not applicable in the present case since $\Gamma_{O}(t)$ is not positive in the entire domain of integration. 
In the inset of Fig. \ref{fig8} we sketch the partial sums $I_O(t)=\sum_{t'=0}^{t}\Gamma_{O}(t')$ for the three algorithms.  While in the SW case the sum $I_O(t)$ rapidly saturates, for the EIC algorithm these sums tend to vanish at infinity and exhibit maxima at the point where the $\Gamma_{O}(t)$ turns negative. We thus consider instead the maxima of these partial sums. Under certain approximation, if the correlation time is much smaller than the time interval $N_a/2$, they may be understood as characteristic times $\tau_{O,b}$. Namely, if $\tau \ll N_a/2$, the exponential decay of autocorrelation functions dies out before the regime of anticorrelations sets in. The maxima of partial sums will then approximate well the (integrated) correlation time, while successfully leaving out of consideration the regime of anticorrelations. 
 Our analysis shows that these maxima (see Fig. \ref{fig10})
 exhibit indeed a power-law scaling with an exponent very close to $z_{O,a}$.
However, if the condition $\tau \ll N_a/2$ is not fulfilled, such an approximation is not justified and we shall discuss the consequences later on. Both $\tau_{O,a}(L)$ and $\tau_{O,b}(L)$ were calculated for the autocorrelations of energy and magnetization and are presented in log-log plots versus the system size $L$ in Fig, \ref{fig11}, 
while the corresponding scaling exponents can be compared in Table III.
\begin{table}[!h]
\begin{ruledtabular}
\caption{Dynamical exponents of EIC algorithm (with $N_a=100$, $v=\frac{1}{10}L^{-\frac{d}{2}}$)}
\centering{
\begin{tabular*}{8.0cm}{c@{}c@{}c@{}}
 & $q=2$, 3D & $q=3$, 2D\\
\hline
\hline
EIC: $z_{m,a}$ & $0.41(5)$ & $0.38(2)$ \\
EIC: $z_{m,b}$ & $0.38(3)$ & $0.29(2)$ \\
EIC: $z_{e,a}$ & $0.42(3)$ & $0.42(2)$ \\
EIC: $z_{e,b}$ & $0.47(3)$ & $0.36(2)$ \\
\hline
SW $z$\footnotemark[2] & $0.46(3)$ & $0.49(1)$ \\
$\alpha/\nu$ & $0.172(1)$\footnotemark[1] & $2/5$ \\
\end{tabular*}
}\end{ruledtabular}
\footnotetext[1]{Best known values \cite{Pelissetto}.}
\footnotetext[2]{Results cited in \cite{OssolaSokal04},\cite{DGMOPS07}}
\label{tab3}
\end{table}

Dynamical exponents derived from the energy and from the magnetization coincide up to the error margins. We also observe a good matching between exponents $z_{O,a}$ and  $z_{O,b}$ in the 3D case, while some discrepancy exists between these exponents in 2D. The reason for that can be related to the choice of the auxiliary parameters.    
\begin{figure}[!hbt]
\includegraphics[scale=0.9]{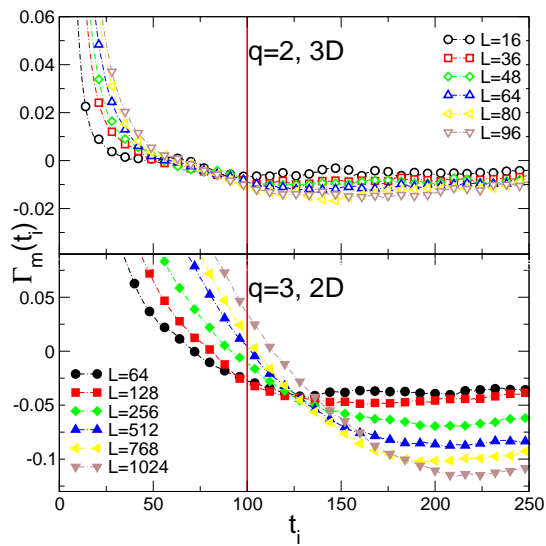}
\caption{(Color online) Behavior of the autocorrelation functions of magnetization for $t_i$ around $N_a=100$ (indicated by the vertical line) in the two cases considered and for various lattice sizes.}
\label{fig12}
\end{figure}
 While the correlation times are several times larger in the 2D case,  all the calculations presented in Table III were performed with the same $N_a=100$. As it may be observed in  Fig. \ref{fig12}, for the 3D Ising case this choice is large enough to fulfill the condition $\tau \ll N_a/2$, and allow the autocorrelations to become negligible before the effect of anticorrelations appears. In 2D case this condition is clearly not fulfilled, especially for larger lattice sizes, and the values $\tau_{m,b}$ and the corresponding exponent cannot be related to the correlation time. It is interesting to discuss these results in context of the argument by Li and Sokal \cite{LiSokal}, which sets the lower bound on the dynamical exponent $z$ in cluster algorithms. Relating the minimal autocorrelation times to the heat capacity of the system, they argue that the lower bound for the dynamical exponent is given by
\begin{equation}
\label{lwbound}
z \geq \alpha/\nu.
\end{equation}   
 The results for $z|_{EIC}$ summarized in Tab. III  are compared to the most recent values of $z|_{SW}$  \cite{OssolaSokal04},\cite{DGMOPS07}, and to the exponent ratio $\alpha/\nu$. In both considered cases the dynamical exponent appears to be comparable, but lower than the value for the SW algorithm. We also observe that the values $\tau_{O,a}$ give results for the dynamical exponent conform to the Li - Sokal bound up to the statistical error. In the 2D case, where discrepancy between the exponents $z_{O,a}$ and  $z_{O,b}$ was observed, the dynamic exponent is just about crossing the Li-Sokal limit, which can be understood as a sign that the system is close to leaving the equilibrium and taking the crossover to IC behavior. It has to be mentioned however, that as long as the width of temperature fluctuations is properly constrained this only seems to increase the noise in our results. No systematic errors were observed in the 2D results for $N_a=100$, while the resulting ensemble retains the correct scaling, proper to the canonical ensemble as illustrated in Figure \ref {fig1}. 

In view of the above discussion we can drive some general conclusions about the choice of the auxiliary parameters $N_a$ and $v$, and their impact on the results.

\subsection{Parameter $N_a$}

\begin{figure}[!hbt]
\includegraphics[scale=0.9]{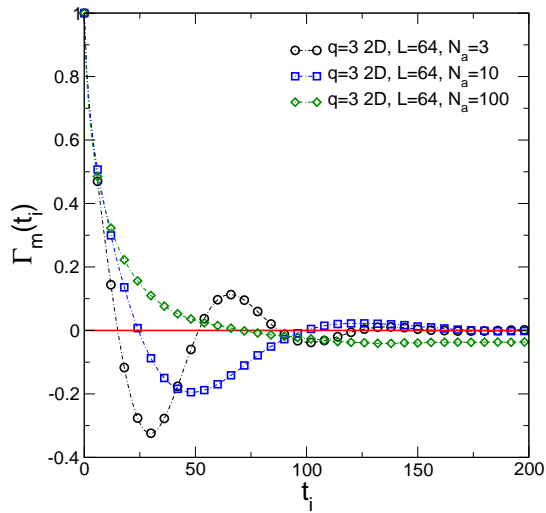}
\caption{(Color online)  Autocorrelation functions of magnetization in the case 2D, $q=3$ for different values of $N_a$.}
\label{fig13}
\end{figure}

   It was briefly mentioned in the end of Sect. \ref{meth} that the number $N_a$ must be large enough to prevent the IC dynamics from taking over the behavior of $p_{\{\sigma\}}$ fluctuations. This is the only restriction to the parameter $N_a$. As discussed earlier in this Section, this starts to happen when $N_a$ is comparable to the value of the autocorrelation time of the system in consideration. Thus $N_a$ should be chosen to be at least more than two times larger than the autocorrelation time for any lattice size considered for a given system. When this is not fulfilled, there is a clear indication of the crossover to the IC dynamics in the behavior of the autocorrelation functions, as observed already in the 2D case with $N_a=100$. When $N_a$ is reduced even further, the autocorrelation function changes the sign more than once, meaning that more than two values of $\overline{p}_{i}$ are correlated, which clearly indicates the domination of IC dynamics over the EIC dynamics (Fig. \ref{fig13}). 
Indeed, in these cases one finds that $\sqrt{var(p_{\{\sigma\}})}\propto L^{-b}$ where $b<\frac{d}{2}$ and the initial requirement for fluctuations of $p_{\{\sigma\}}$ is not fulfilled. 

\subsection{Parameter $v$}

   The effects of varying the parameter $v$ can also be observed in the behavior of autocorrelation functions. We shall illustrate this on two examples. In the first example $v$ was augmented by keeping the the power-law decay of the form $L^{-d/2}$, but increasing the constant of proportionality $\tilde v$ beyond the value compatible with the width following from the binomial distribution of Eq. \ref{fixed-s}. In the second one, the parameter $v$ lies inside the regime imposed by the EIC constraint, but decreases with size faster than $L^{-d/2}$. 

The examples are taken in the 2D, $q=3$ case for the lattice sizes from $L=64$ to $640$, and using statistics of $10^6$ MC steps, while $N_a$  was increased to $N_a=250$ in order to avoid its contribution to nonequilibrium effects. For the case 2D, $q=3$, the width of the distribution of $p_{\{\sigma\}}$ corresponding to the binomial distribution in Eq. (\ref{fixed-s}) would be $\approx \frac{\sqrt{p_C(1-p_C)}}{2}\cdot n_{ss}^{-\frac{d}{2}}\approx 0.19\cdot L^{-1}$. We compare the results for the three different values of $v$: (a) $v=\frac{1}{3}\cdot L^{-1}$;  (b) $v=\frac{1}{10}\cdot L^{-1}$ and (c) $v=1\cdot L^{-1.5}$. The results for the dynamical exponents are shown in the Tab. IV.
\begin{table}[!h]
\begin{ruledtabular}
\caption{Dynamical exponents for the case 2D, $q=3$ depending on $v$ (with $N_a=250$)}
\centering{
\begin{tabular*}{8.0cm}{c@{}c@{}c@{}c@{}}
 & $v=\frac{1}{3}\cdot L^{-1}$ & $v=\frac{1}{10}\cdot L^{-1}$ & $v=1\cdot L^{-1.5}$ \\
\hline
\hline
EIC: $z_{m,a}$ & - & $0.39(3)$ & $0.56(2)$ \\ 
EIC: $z_{m,b}$ & $0.20(3)$ & $0.34(4)$ & $0.53(2)$  \\
EIC: $z_{e,a}$ & - & $0.41(3)$ & $0.52(2)$ \\
EIC: $z_{e,b}$ & $0.33(4)$ & $0.38(4)$ & $0.54(2)$ \\
\end{tabular*}
}\end{ruledtabular}
\label{tab4}
\end{table}

Let us first comment the reference case (b) corresponding to the value of $v$ used throughout this paper. When comparing the results with those in Table IV, obtained with $N_a=100$, one may see that  by increasing $N_a$ the exponent $z_{O,a}$ approaches closer to the Li-Sokal limit, while the discrepancy between exponents $z_{O,a}$ and $z_{O,b}$ diminishes. This confirms the earlier discussion in this Section and the expectation that an increasing of $N_a$ would indeed correct the deviation from the equilibrium behavior.

In the case (a) the width $v$ has the same power-law decay,  but the constant factor $\tilde v$ was taken larger than the value used in this paper and also larger than the one of the binomial distribution in Eq. (\ref{fixed-s}). In spite of the correct power-law dependence, the crossover to the IC regime for lattice sizes considered here is observed. In autocorrelation functions, the regime of single exponential decay shrinks drastically and consequently times $\tau_{O,a}$ could not be defined. We may still consider the quantity $\tau_{O,b}$, since it still exhibits a power-law behavior, although it may not be interpreted as a correlation time. The corresponding exponent $z_{O,b}$ is much smaller than the Li-Sokal limit for the dynamical exponent of the equilibrium cluster algorithms.

In contrast to this, in the example (c) with $v=1\cdot L^{-1.5}$ the exponential regime of the autocorrelation functions is longer then for $v=\frac{1}{10}\cdot L^{-1}$ and consequently allows a more accurate calculation of both $\tau_{O,a}$ and $\tau_{O,b}$, giving the same dynamical exponent up to the error margins. It is interesting to notice that by reducing the width parameter $v$, the dynamic exponent $z$ has increased. This provides a tool to vary the dynamical exponent continuously, by varying the exponent of power-law decay in $v$.

   
\section{Discussion and conclusion\label{discs}}

In this paper we presented more detailed study of recently proposed EIC algorithm, designed to simultaneously drive the system to criticality without its prior knowledge and to provide the correct critical exponents. Extensive numerical calculations performed on two typical examples of the Potts model in two and three dimensions confirmed that the method efficiently produces results for both critical exponents and critical temperature, with considerable accuracy that  attains up to 4 digits for critical exponents and six digits in the case of critical temperature. We conclude that the two auxiliary parameters of the algorithm, $N_a$ and $v$, may be varied in the wide range without producing systematic errors, and that their influence on the results when exceeding this range can be controlled by examining the behavior of the autocorrelation functions. We have also estimated the dynamic exponent of the EIC algorithm, and shown that it can be tuned by varying the auxiliary parameters. It is found to be larger than the one of the standard IC algorithm and generally smaller than the exponent of the SW algorithm. We also notice that, when the dynamical exponent is lowered and reaches beyond the Li-Sokal limit, the crossover to the nonequilibrium IC algorithm can be observed.

We observe that both in the SW method and in the present EIC calculations, the dynamical exponent does not change significantly with dimensionality and stays in the range between 0.4 and 0.5 in both cases, while the the Li-Sokal limit differs by more than twice between the two cases. The fact that within the EIC algorithm the dynamical exponent $z$ may be tuned by changing the auxiliary parameters provides the means to make further analysis in order to better explain the reasons, still not well understood  \cite{DGMOPS07}, why the equilibrium cluster algorithms like SW in some cases (e.g. for 3D Ising), have much larger dynamical exponent $z$ than required by the Li-Sokal limit. 

Due to its capability of self-tuning to the critical point there is a wide range of possible applications of the EIC algorithm to phase transitions, where the critical temperature is not known in advance. One of the most appealing is certainly for simulations of phase transitions in presence of quenched disorder, where it provides the means to calculate critical properties at finite size critical temperature for each individual configuration at feasible time, which helps to overcome the problems related to the lack of self-averaging \cite{BUprep}. 

The algorithm could also be generalized  to other, or more complex problems such as the first order phase transition only briefly mentioned here, or to tricritical points, for which the extension of the standard IC algorithm was already applied \cite{BU07}.

\begin{acknowledgements}

This work was supported by the Croatian Ministry of Science, Education and Sports through grant No. 035-0000000-3187.

\end{acknowledgements}                       
                                                                              
\vfill

\end{document}